\documentstyle[12pt,aps,tighten]{revtex}
\begin{document}
\draft
\title{Transmission spectrum of a tunneling particle interacting with\\
dynamical fields: real-time functional-integral approach}
\author{Masahito Ueda}
\address{Department of Physical Electronics, Hiroshima University, Higashi-Hiroshima 739, Japan}
\date{\today}
\maketitle
\begin{abstract}
A real-time functional-integral method is used to derive an effective action that gives the transmission spectrum of a tunneling particle interacting with a bath of harmonic oscillators.
The transmission spectum is expressed in terms of double functional integrals with respect to the coordinate of the particle which are evaluated by means of stationary-phase approximation.
The equations of motion for the stationary-phase trajectories are solved exactly for an arbitrary spectral density function of the bath, and the obtained solutions are used to find the transmission spectra for specific examples.
For a bath with single frequency $\omega$, an analytic expression of the transmission spectrum is obtained which covers from sudden tunneling ($\omega T_0\ll 1$) to adiabatic one ($\omega T_0\gg 1$),
where $T_0$ is the time it would take a classical particle to traverse the inverted bare potential barrier.
For a bath with Ohmic spectrum, the differential tunneling conductance at low bias voltage $V$ and for $\eta T_0\ll 1$ is found to obey a power law $\sim(eVT_0/\hbar)^{\eta T_0S_0/2\pi\hbar}$, where $\eta$ is the friction coefficient and $S_0$ is the tunneling exponent in the absence of interaction.
\end{abstract}
\pacs{PACS numbers: 73.40.Gk}


\section{Introduction}
\label{sec:Introduction}
Persistent efforts have been devoted to the development of computational methods in tunneling.
The instanton method offers a way of evaluating the decay rate of a metastable state~\cite{Langer,Coleman}.
The central idea of this method is to deduce the information on the lifetime of a metastable state from the imaginary part of its free energy that could only be defined by means of analytic continuation.
The instanton method, however, does not give the transmission spectrum ---the energy distribution of transmitted particles--- because the free energy is calculated from the partition function which is defined without reference to the final states of the transmitted particles.
A modified version of the instanton method which takes only half of the closed-loop bounce trajectory can compute the tunneling rate for the case in which both initial and final states of the bath are in the ground state~\cite{Sebastian}, but the instanton method has not succeeded in finding full transmission spectrum~\cite{Bruinsma}.

Quite often, however, one encounters situations in which one needs to find the transmission spectrum.
Examples are the Coulomb blockade of tunneling~\cite{Grabert},
             deep inelastic collisions of heavy ions~\cite{Mohring},
             atom-surface scattering~\cite{Newns},
             chemical reactions~\cite{Miller}, and, of course, 
             tunneling spectroscopy~\cite{Wolf}.
Ueda and Ando have recently computed full transmission spectrum of a tunneling particle interacting with a bath of harmonic oscillators by combining an operator-algebraic method and a functional-integral one followed by analytic continuation to imaginary time~\cite{UA1}.
Real-time approaches have also been used to find information on transmission spectrum.
M\"{o}hring and Smilansky~\cite{Mohring} used the influence functional~\cite{Feynman} to calculate the variance of the energy that a tunneling particle exchanges with the bath.
By applying a time-dependent WKB method to approximate functional integrals for the coordinate of the particle and then by performing functional integrals for the bath degrees of freedom, Bruinsma and Bak derived the effective action to the lowest order in the coupling constant~\cite{Bruinsma}.

The primary purpose of this paper is to use a real-time functinal-integral method to derive an {\it exact} expression of the transmission spectrum of a tunneling particle interacting with a bath of harmonic oscillators,
and to show how to get information on the tunneling exponent from the exact expression by means of stationary-phase approximation.
The method presented in this paper allows a systematic expansion of the effective action in terms of the coupling constant and takes account of the fact that the time $T$ it would take the tunneling particle to traverse the barrier along the stationary-phase trajectory is modifed by the interaction
--- a point that has eluded the treatment using a time-dependent WKB approximation~\cite{Bruinsma}.
The present method also allows us to analytically study the dynamics of tunneling from the sudden limit $\omega T_0\ll 1$ to the adiabatic one $\omega T_0\gg 1$,
where $\omega$ is the characteristic frequency of the bath and $T_0$ is the time it would take a classical particle to traverse the inverted bare potential barrier.

This paper is organized as follows.
Section~\ref{sec:Formulas} obtains a general formula for the transmission spectrum (Sec.~\ref{sec:General T}) and applies it to the Caldeira-Leggett model (Sec.~\ref{sec:Effective})~\cite{CL}, where the transmission spectrum is expressed in terms of double functional integrals with respect to the coordinate of the particle alone.
Section~\ref{sec:Stationary} evaluates the double functional integrals by means of stationary-phase approximation and exactly solves the equations of motion for the stationary-phase trajectories for an arbitrary spectral density function of the bath.
The obtained solutions are used to analyze the case of a single-frequency bath in Sec.~\ref{sec:Single} and the case of an Ohmic bath in Sec.~\ref{sec:Ohmic}.
The relation of the present method to the instanton method is discussed in Sec.~\ref{sec:Instanton}.
Section~\ref{sec:bltime} examines the distribution of sideband intensities of particles transmitting through a quantum-mechanically fluctuating barrier and compares the results with those obtained for the case in which the barrier is time-modulated in a prescribed manner~\cite{BL}.

\section{Formula for the transmission spectrum}
\label{sec:Formulas}
\subsection{General theory}
\label{sec:General T}

We consider a situation in which a particle of mass $M$ tunnels through a potential barrier $V(X)$ while interacting with other degrees of freedom -- ^^ ^^ the bath."
A general scattering-theory formula for the rate of transition from state  $|\Psi_{\rm i}\rangle$ to state  $|\Psi_{\rm f}\rangle$ is given in the Schr\"{o}dinger representation by
\begin{eqnarray}
\Gamma_{{\rm i}\rightarrow{\rm f}}=\!\!\lim_{t_{\rm f,i}\rightarrow\pm\infty}
\!\frac{1}{t_{\rm f}-t_{\rm i}}|\langle\Psi_{\rm f}|
T\exp\!\left(-\frac{i}{\hbar}\int^{t_{\rm f}}_{t_{\rm i}}\!Hdt\right)
|\Psi_{\rm i}\rangle|^2,
\label{2A.4}
\end{eqnarray}
where $H$ is the total Hamiltonian and $T$ is the time-ordering operator.
When $H$ does not depend on time explicitly, the factor
$T\exp\!\left(-\frac{i}{\hbar}\int^{t_{\rm f}}_{t_{\rm i}}\!Hdt\right)$
may be replaced by $e^{-iH(t_{\rm f}-t_{\rm i})/\hbar}$.
When we are not interested in the state of the bath, we make a statistical average of Eq.~(\ref{2A.4}) over initial states of the bath and sum over final states:
\begin{eqnarray}
\Gamma(E,E')
=\!\!\!\lim_{t_{\rm f,i}\rightarrow\pm\infty}\frac{1}{t_{\rm f}-t_{\rm i}}
\sum_{\rm i,f}P(\epsilon^{\rm i}_{\rm b})
|\langle\Psi_{\rm f}|
T\exp\!\left(-\frac{i}{\hbar}\int^{t_{\rm f}}_{t_{\rm i}}\!Hdt\right)
|\Psi_{\rm i}\rangle|^2
\delta[E-E'-(\epsilon^{\rm f}_{\rm b}-\epsilon^{\rm i}_{\rm b})],
\label{2A.5}
\end{eqnarray}
where $E$ is the initial-state energy of the particle, $\epsilon^{\rm i}_{\rm b}$ and $\epsilon^{\rm f}_{\rm b}$ are the initial-state and final-state energies of the bath, and $P(\epsilon^{\rm i}_{\rm b})$ denotes the probability distribution of $\epsilon^{\rm i}_{\rm b}$.
The delta function restricts the energy gain (if $E-E'>0$ or loss otherwise) of the bath between the initial state and the final state to be equal to $E-E'$ which at the limit of $t_{\rm f}-t_{\rm i}=\infty$ becomes the energy loss ( if $E-E'>0$ or gain otherwise) of the tunneling particle.
Therefore $E'$ has the meaning of the final-state energy of the particle and
Eq.~(\ref{2A.5}) may be written as
\begin{eqnarray}
\Gamma(E,E')
=\!\!\!\lim_{t_{\rm f,i}\rightarrow\pm\infty}\frac{1}{t_{\rm f}-t_{\rm i}}
\sum_{\rm i,f}P(\epsilon^{\rm i}_{\rm b})
|\langle E',\epsilon^{\rm f}_{\rm b}|
T\exp\!\left(-\frac{i}{\hbar}\int^{t_{\rm f}}_{t_{\rm i}}\!Hdt\right)
| E,\epsilon^{\rm i}_{\rm b}\rangle|^2
\delta[E-E'-(\epsilon^{\rm f}_{\rm b}-\epsilon^{\rm i}_{\rm b})].
\label{2A.5'}
\end{eqnarray}
Inserting into Eq.~(\ref{2A.5'}) the completeness relations for the coordinate of the particle in the initial and final states, we obtain
\begin{eqnarray}
\Gamma(E,E')
&=&\!\lim_{t_{\rm f,i}\rightarrow\pm\infty}\frac{1}{t_{\rm f}-t_{\rm i}}
\!\int^\infty_{-\infty}\!\frac{dt}{2\pi\hbar}e^{\frac{i}{\hbar}(E-E')t}
\!\!\int\!\!dX_{\rm f}\!\!\int\!\!dX_{\rm i}\!\!\int\!\!dY_{\rm f}\!\!\int\!\!dY_{\rm i} \nonumber\\
& & \times
\langle E'|X_{\rm f} \rangle\langle  X_{\rm i}|E  \rangle
\langle E|Y_{\rm i}  \rangle\langle  Y_{\rm f}|E' \rangle
A(X_{\rm f},X_{\rm i},Y_{\rm f},Y_{\rm i};t),
\label{2A.6}
\end{eqnarray}
where  
\begin{eqnarray}
\!\!A(X_{\rm f},X_{\rm i},Y_{\rm f},Y_{\rm i};t)
&=&\sum_{\rm i,f}P(\epsilon^{\rm i}_{\rm b})
\langle X_{\rm f},\epsilon^{\rm f}_{\rm b}|
T\exp\!\left(-\frac{i}{\hbar}\int^{t_{\rm f}}_{t_{\rm i}}\!Hdt\right)
|X_{\rm i},\epsilon^{\rm i}_{\rm b}\rangle
\nonumber \\
& & \times
\langle Y_{\rm i},\epsilon^{\rm i}_{\rm b}|
\tilde{T}\exp\!\left(\frac{i}{\hbar}\int^{t_{\rm f}}_{t_{\rm i}}\!Hdt\right)
|Y_{\rm f},\epsilon^{\rm f}_{\rm b}\rangle \
e^{-\frac{i}{\hbar}(\epsilon^{\rm f}_{\rm b}-\epsilon^{\rm i}_{\rm b})t}\!,
\label{2A.7}
\end{eqnarray}
where $\tilde{T}$ is the counter-time-ordering operator.
Equation (\ref{2A.6}) together with Eq.~(\ref{2A.7}) gives the transmission spectrum that a particle with initial-state energy $E$ transmits through a barrier with final-state energy $E'$.
This formula applies whether or not the situation of our concern is related to tunneling. 
The condition that the particle indeed tunnels through the barrier will be imposed later [Eq.~(\ref{3A.8})].

\subsection{Effective action for the Caldeira-Leggett model}
\label{sec:Effective}

By explicitly evaluating Eq.~(\ref{2A.7}) for the Caldeira-Leggett model~\cite{CL}, we derive an {\it exact} effective action for the transmission spectrum of a particle interacting with a bath of harmonic oscillators.
The Hamiltonian of the Caldeira-Leggett model is given by
\begin{eqnarray}
H=\frac{P^2}{2M}+V(X)
+\sum_\alpha\left\{\frac{p_\alpha^2}{2m_\alpha}
+\frac{m_\alpha\omega_\alpha^2}{2}\left[x_\alpha-f_\alpha(X)\right]^2\right\},
\label{2B.1}
\end{eqnarray}
where $X$ and $P$ are the coordinate and momentum of the particle;
$x_\alpha$ and $p_\alpha$ are the canonically-conjugate generalized coordinate and momentum of mode $\alpha$ of the bath, and $m_\alpha$ and $\omega_\alpha$ are its mass and frequency; and $f_\alpha(X)$ is a function that characterizes the interaction between the particle and the bath.

The Caldeira-Leggett model includes the counter term $\sum_\alpha(m_\alpha\omega_\alpha^2/2)f^2_\alpha(X)$.
Whether or not this term should be included depends on the problem concerned, but we proceed with it as it can easily be subtracted at any stage in the following discussions.
Because the transmission spectrum of a particle is meaningful only if the particle is free in the initial and final states, we require that
\begin{eqnarray}
V(X_{\rm i(f)})=V(Y_{\rm i(f)})=0
\label{assumption1} 
\end{eqnarray}
and
\begin{eqnarray}
f_\alpha(X_{\rm i(f)})=f_\alpha(Y_{\rm i(f)})\equiv f_\alpha^{\rm i(f)},
\label{assumption2}
\end{eqnarray}
where $f_\alpha^{\rm i(f)}$ may depend on parameters such as $m_\alpha$, $\omega_\alpha$, etc., but should not depend on the dynamical variables of the particle.
If these requirements are met, the particle and the bath are decoupled in the initial and final states. The energy eigenstate of the bath in the initial (final) state $|\epsilon^{\rm i(f)}_{\rm b}\rangle$ is represented by $|\{m^{\rm i(f)}_\alpha\}\rangle$ with its eigenenergy given by
\begin{eqnarray}
\epsilon^{\rm i(f)}_{\rm b}
=\!\sum_{\{m^{\rm i(f)}_\alpha\}}\!\hbar\omega_\alpha\!
\left(m^{\rm i(f)}_\alpha+\frac{1}{2}\right),
\label{2B.2}
\end{eqnarray}
where $m^{\rm i(f)}_\alpha$ denotes the excitation number of mode $\alpha$ of the bath.

To eliminate the bath degrees of freedom, we express Eq.~(\ref{2A.7}) in coordinate representation of the bath by inserting the completeness relations:
\begin{eqnarray}
& & A(X_{\rm f},X_{\rm i},Y_{\rm f},Y_{\rm i};t)=
\prod_\alpha
\int^\infty_{-\infty} \!\!\!dx_{\alpha1}\int^\infty_{-\infty} \!\!dx_{\alpha2}\int^\infty_{-\infty} \!\!dx_{\alpha3}\int^\infty_{-\infty} \!\!dx_{\alpha4} 
\nonumber \\
& & \times
\langle X_{\rm f},\{x_{\alpha 1}\}|e^{-\frac{i}{\hbar}H(t_{\rm f}-t_{\rm i})}
|X_{\rm i},\{x_{\alpha 2}\} \rangle
\langle Y_{\rm i},\{x_{\alpha 3}\}|e^{ \frac{i}{\hbar}H(t_{\rm f}-t_{\rm i})}
|Y_{\rm f},\{x_{\alpha 4}\} \rangle
\nonumber \\
& & \times 
\sum_{m_\alpha^{\rm i}}\!
P(m_\alpha^{\rm i})
\langle x_{\alpha 2}|m_\alpha^{\rm i} \rangle
\langle m_\alpha^{\rm i}|x_{\alpha 3} \rangle e^{i\omega_\alpha\left(m_\alpha^{\rm i}+\frac{1}{2}\right) t}
\sum_{n_\alpha^{\rm f}}
\langle n_\alpha^{\rm f}|x_{\alpha 1} \rangle
\langle x_{\alpha 4}|n_\alpha^{\rm f} \rangle e^{-i\omega_\alpha \left(n_\alpha^{\rm f}+\frac{1}{2}\right)t}.
\label{2B.4}
\end{eqnarray}
Here $\langle x|m_\alpha^{\rm i}\rangle$ and $\langle x|n_\alpha^{\rm f}\rangle$ are given in terms of the wavefunction $\psi_n(x)$ of the harmonic oscillator~\cite{Landau} as $\psi_{m_\alpha}\!\left(x-f_\alpha^{\rm i}\right)$ and $\psi_{n_\alpha}\!\left(x-f_\alpha^{\rm f}\right)$, respectively.
The summation over $n_\alpha^{\rm f}$ gives the Green's function of the shifted harmonic oscillator:
\begin{eqnarray}
\sum_{n_\alpha^{\rm f}}\cdots
=\sqrt{\frac{m_\alpha\omega_\alpha}{2\pi i\hbar\sin\omega_\alpha t}}
\exp\!\left\{\!
\frac{
m_\alpha\omega_\alpha\!\left[
2\tilde{x}_{\alpha 1}\tilde{x}_{\alpha 4}-\left(\tilde{x}_{\alpha 1}^2+\tilde{x}_{\alpha 4}^2\right)\cos\omega_\alpha t\right]}
{2i\hbar\sin\omega_\alpha t}
\!\right\},
\label{2B.5}
\end{eqnarray}
where $\tilde{x}_{\alpha 1(4)}\equiv x_{\alpha 1(4)}-f_\alpha^{\rm f}$.
When the initial state of the bath is at thermal equilibrium with
temperature $\beta^{-1}$, $P(m_\alpha^{\rm i})$ is given by
\begin{eqnarray}
P(m_\alpha^{\rm i})=(1-e^{-\beta\hbar\omega_\alpha})
e^{-\beta m_\alpha^{\rm i}\hbar\omega_\alpha},
\label{2B.6}
\end{eqnarray}
and the summation over $m_\alpha^{\rm i}$ can be carried out to yield
\begin{eqnarray}
\sum_{m_\alpha^{\rm i}}\cdots
&=&2\sinh\!\frac{\beta\hbar\omega_\alpha}{2}
\sqrt{\frac{m_\alpha\omega_\alpha}
{-2i\pi\hbar\sin\omega_\alpha(t+i\beta\hbar)}}
\nonumber \\
& & \times
\exp\!\left\{\!
\frac{m_\alpha\omega_\alpha\!\left[2\tilde{x}_{\alpha 2}\tilde{x}_{\alpha 3}\!-\!\left(\tilde{x}_{\alpha 2}^2\!+\!\tilde{x}_{\alpha 3}^2\right)\cos\omega_\alpha(t\!+\!i\beta\hbar)\right]}{-2i\hbar\sin\omega_\alpha(t+i\beta\hbar)}
\!\right\}\!,
\label{2B.7}
\end{eqnarray}
where $\tilde{x}_{\alpha 2(3)}\equiv x_{\alpha 2(3)}-f_\alpha^{\rm i}$.
The probability amplitudes $\langle X_{\rm f},\{x_{\alpha 1}\}|e^{-\frac{i}{\hbar}H(t_{\rm f}-t_{\rm i})}|X_{\rm i},\{x_{\alpha 2}\} \rangle$ and
$\langle X_{\rm i},\{x_{\alpha 3}\}|e^{\frac{i}{\hbar}H(t_{\rm f}-t_{\rm i})}|X_{\rm f},\{x_{\alpha 4}\} \rangle$
may be expressed in terms of functional integrals over the coordinate of the particle and over the coordinates of the bath.
Because the Hamiltonian is quadratic with respect to the coordinates of the bath and the interaction term is linear in them, path integrations over the bath degrees of freedom can be carried out explicitly, giving~\cite{Feynman}
\begin{eqnarray}
& & \langle X_{\rm f},\{x_{\alpha 1}\}|e^{-\frac{i}{\hbar}H(t_{\rm f}-t_{\rm i})}|X_{\rm i},\{x_{\alpha 2}\}\rangle 
= \prod_\alpha
\sqrt{\frac{m_\alpha\omega_\alpha}{2\pi i\hbar\sin\omega_\alpha (t_{\rm f}-t_{\rm i})}} 
\nonumber \\
& & \times
\int \!\!\!\!\! \int^{X(t_{\rm f})=X_{\rm f}}_{X(t_{\rm i})=X_{\rm i}}
\!\!{\cal D}X
\exp\!\left\{\!
\frac{i}{\hbar}\!\!\int^{t_{\rm f}}_{t_{\rm i}}\!\!\!\!dt_1\!
\left[\frac{M}{2}\dot{X}^2(t_1)\!-\!V[X(t_1)]\!
-\!\frac{m_\alpha\omega_\alpha^2}{2}f_\alpha^2[X(t_1)]\right]
+\sum_\alpha\frac{im_\alpha\omega_\alpha}{2\hbar\sin\omega_\alpha (t_{\rm f}-t_{\rm i})}
\right.
\nonumber \\
& & 
\times
\!\left[(x_{\alpha 1}^2\!+\!x_{\alpha 2}^2)\cos\omega_\alpha 
(t_{\rm f}\!-\!t_{\rm i})\!-\!2x_{\alpha 1}x_{\alpha 2}+2\omega_\alpha\!\int^{t_{\rm f}}_{t_{\rm i}}\!\!\!dt_1 f_\alpha[X(t_1)]
\left[x_{\alpha 1}\sin\omega_\alpha \!(t_1\!-\!t_{\rm i})\!+\!x_{\alpha 2}\sin\omega_\alpha(t_{\rm f}\!-\!t_1)\right]\right.
\nonumber \\
& &
\left.\left.
-2\omega_\alpha^2\!\!\int^{t_{\rm f}}_{t_{\rm i}}\!\!\!\!dt_1\!\!\!
\int^{t_{\rm f}}_{t_{\rm i}}\!\!\!\!dt_2
f_\alpha[X(t_1)]f_\alpha[X(t_2)]\sin\omega_\alpha(t_{\rm f}\!-\!t_1)\sin\omega_\alpha (t_2\!-\!t_{\rm i})
\right]\!\right\}\!,\nonumber \\
\label{2B.8}
\end{eqnarray}
and a similar expression can be found for $\langle Y_{\rm i},\{x_{\alpha 3}\}|e^{\frac{i}{\hbar}H(t_{\rm f}-t_{\rm i})}|Y_{\rm f},\{x_{\alpha 4}\}\rangle$.
Substituting Eqs.~(\ref{2B.5}), (\ref{2B.7}) and (\ref{2B.8}) into Eq.~(\ref{2B.4}) and performing integration over $x_{\alpha i}$ ($i=1,\cdots, 4$) finally 
yield
\begin{eqnarray}
A(X_{\rm f},X_{\rm i},Y_{\rm f},Y_{\rm i};t) = \!
\int \!\!\!\!\! \int^{X(t_{\rm f})=X_{\rm f}}_{X(t_{\rm i})=X_{\rm i}}
\!\! {\cal D}X \!
\int \!\!\!\!\! \int^{Y(t_{\rm f})=Y_{\rm f}}_{Y(t_{\rm i})=Y_{\rm i}}
\!\! {\cal D}Y  
\exp\!\left[\frac{i}{\hbar}S_{\rm eff}(t)\right]\!,
\label{2B.9}
\end{eqnarray}
where the effective action $S_{\rm eff}(t)$ is given in matrix form as
\begin{eqnarray}
& & S_{\rm eff}(t)=\int^{t_{\rm f}}_{t_{\rm i}}\!\!\!dt_1
\left[\frac{M}{2}\dot{X}^2-\frac{M}{2}\dot{Y}^2-V(X)+V(Y)\right]
+\sum_\alpha 
\frac{im_\alpha\omega_\alpha}{4\sinh\frac{\beta\hbar\omega_\alpha}{2}}
\!\int^{t_{\rm f}}_{t_{\rm i}}\!\!\!dt_1\int^{t_{\rm f}}_{t_{\rm i}}\!\!\!dt_2
\nonumber \\
& & \times
\left[\!
      \begin{array}{c}
        \dot{f}_\alpha[X(t_1)],-\dot{f}_\alpha[Y(t_1)]
      \end{array}
\!\right]
\!\!\left[\!
\begin{array}{cc}
\!\cosh\!\omega_\alpha\!\!\left(\!\frac{\beta\hbar}{2}-i|t_1\!-\!t_2|\!\right) 
& \!\!\!\!\cosh\!\omega_\alpha\!\!\left(\!\frac{\beta\hbar}{2}+i(t_1\!-\!t_2\!-\!t)\!\right)\!\!
\\
\!\cosh\!\omega_\alpha\!\!\left(\!\frac{\beta\hbar}{2}-i(t_1\!-\!t_2\!+\!t)\!\right) 
& \!\!\!\!\cosh\!\omega_\alpha\!\!\left(\!\frac{\beta\hbar}{2}+i|t_1\!-\!t_2|\!\right)\!\!\!
\end{array}
\!\right]
\!\left[
\begin{array}{r}
\!\dot{f}_\alpha[X(t_2)]\! \\
\!-\dot{f}_\alpha[Y(t_2)]\!
\end{array}
\right].
\label{2B.10}
\end{eqnarray}
This is the desired effective action for the Caldeira-Leggett model that gives the transmission spectrum of a tunneling particle interacting with a bath of harmonic oscillators.

For the case of a separable interaction~\cite{CL} in which $f_\alpha(X)$ is given by
$f_\alpha(X)=C_\alpha g(X)/m_\alpha\omega_\alpha^2$,
Eq.~(\ref{2B.10}) reduces to 
\begin{eqnarray}
S_{\rm eff}(t)\!&=&\!\int^{t_{\rm f}}_{t_{\rm i}}\!\!dt_1
\left[\frac{M}{2}\dot{X}^2-\frac{M}{2}\dot{Y}^2-V(X)+V(Y)\right]\nonumber \\
& &
+ i\!\int^{t_{\rm f}}_{t_{\rm i}}\!\!dt_1\!\int^{t_{\rm f}}_{t_{\rm i}}\!\!dt_2\!
\left[\!
\begin{array}{c}
\dot{g}[X(t_1)],-\dot{g}[Y(t_1)]
\end{array}
\!\right]\!\!\!
\left[
\begin{array}{cc}
\!\alpha^{\rm c}(t_1-t_2) & \!\alpha^<(t_1-t_2-t)\! \\
\!\alpha^>(t_1-t_2+t)     & \!\alpha^{\rm ac}(t_1-t_2)\!
\end{array}
\right]
\!\!\!
\left[
\begin{array}{r}
 \!\dot{g}[X(t_2)]\! \\
\!-\dot{g}[Y(t_2)]\!
\end{array}
\right],
\label{2B.11}
\end{eqnarray}
where the nonlocal integration kernels
\begin{eqnarray}
\alpha^>(t)&=&\left[\alpha^<(t)\right]^*=
\int^\infty_{-\infty}\frac{d\omega}{2\pi}\frac{J(\omega)}{\omega^2}\frac{\cosh\omega\left(\frac{\beta\hbar}{2}-it\right)}{\sinh\frac{\beta\hbar\omega}{2}}, 
\nonumber \\
\alpha^{\rm c}(t)&=&\left[\alpha^{\rm ac}(t)\right]^*=\alpha^>(|t|),
\label{2B.12}
\end{eqnarray}
are characterized by the spectral density function of the bath
\begin{eqnarray}
J(\omega)=\frac{\pi}{2}\sum_\alpha \frac{C_\alpha^2}{m_\alpha\omega_\alpha}
\delta(\omega-\omega_\alpha).
\label{2B.13}
\end{eqnarray}

The Green's function matrix appearing in Eq.~(\ref{2B.11}) reminds us of the closed-time-path Green's function (CTPGF) formalism~\cite{Schwinger}, which pursues the time evolution of the system by eliminating all pieces of information on the bath degrees of freedom.
The present theory, by contrast, does not eliminate all of them but keeps the piece of information on how much energy has been transferred between the system and the bath.
Hence we have the time argument $t$ in the off-diagonal elements of the Green's function matrix in Eq.~(\ref{2B.11}); it plays the role of giving the transmission spectrum after the Fourier transformation in Eq.~(\ref{2A.6}).

Substituting Eq.~(\ref{2B.9}) into Eq.~(\ref{2A.6}) gives
\begin{eqnarray}
\Gamma(E,E')
&=&\!\lim_{t_{\rm f,i}\rightarrow\pm\infty}\frac{1}{t_{\rm f}-t_{\rm i}}
\!\int^\infty_{-\infty}\!\!\frac{dt}{2\pi\hbar}e^{\frac{i}{\hbar}(E-E')t}
\!\!\int \!\!dX_{\rm f}\!\!\int \!\!dX_{\rm i}\!\!\int \!\!dY_{\rm f}\!\!\int \!\!dY_{\rm i} 
\langle E'|X_{\rm f} \rangle\langle  X_{\rm i}|E  \rangle
\langle E|Y_{\rm i}  \rangle\langle  Y_{\rm f}|E' \rangle
\nonumber\\
& & \times
\int \!\!\!\! \int^{X(t_{\rm f})=X_{\rm f}}_{X(t_{\rm i})=X_{\rm i}} \!\!
{\cal D}X \!\!
\int \!\!\!\! \int^{Y(t_{\rm f})=Y_{\rm f}}_{Y(t_{\rm i})=Y_{\rm i}} \!\!
{\cal D}Y  
\exp\!\left[\frac{i}{\hbar}S_{\rm eff}(t)\right]\!.
\label{2B.14}
\end{eqnarray}
The integrals in Eq.~(\ref{2B.14}) over  $X_{\rm f}$, $X_{\rm i}$, $Y_{\rm f}$, $Y_{\rm i}$ give the boundary conditions on functional integrals~\cite{Newns}.
To see this, let us introduce the center-of-mass coordinate $R\equiv(X+Y)/2$ and the relative coordinate $q\equiv X-Y$.
The relevant part of the effective action  becomes
\begin{eqnarray}
\int^{t_{\rm f}}_{t_{\rm i}}dt_1\left[\frac{M}{2}\dot{X}^2-\frac{M}{2}\dot{Y}^2\right]=
\int^{t_{\rm f}}_{t_{\rm i}}M\dot{R}\dot{q}dt_1
=M\left[
\dot{R}(t_{\rm f})q(t_{\rm f})-\dot{R}(t_{\rm i})q(t_{\rm i})
\right]-\int^{t_{\rm f}}_{t_{\rm i}}M\ddot{R}qdt_1.
\end{eqnarray}
Since there is neither the potential nor the interaction in the initial and final states as required in Eqs.~(\ref{assumption1}) and (\ref{assumption2}), there is no acceleration $\ddot{R}$.
Noting that
\begin{eqnarray}
\langle X_{\rm i}|E\rangle =e^{ik_{\rm i}X_{\rm i}}, \ \
\langle E'|X_{\rm f}\rangle=e^{-ik_{\rm f}X_{\rm f}},
\end{eqnarray}
where
$k_{\rm i}\equiv\sqrt{2ME/\hbar^2}$ and 
$k_{\rm f}\equiv\sqrt{2ME'/\hbar^2}$,
we find that
\begin{eqnarray}
& & \!\!\int \!\!dX_{\rm f}\!\!\int \!\!dX_{\rm i}\!\!\int \!\!dY_{\rm f}\!\!\int \!\!dY_{\rm i} 
\langle E'|X_{\rm f} \rangle\langle  X_{\rm i}|E  \rangle
\langle E|Y_{\rm i}  \rangle\langle  Y_{\rm f}|E' \rangle
e^{\frac{i}{\hbar}\left[M\dot{R}(t_{\rm f})(X_{\rm f}-Y_{\rm f})-
M\dot{R}(t_{\rm i})(X_{\rm i}-Y_{\rm i})\right]}
\nonumber \\
& & 
=(2\pi\hbar)^4\left[
\delta(M\dot{R}_{\rm f}-\hbar k_{\rm f})
\delta(M\dot{R}_{\rm i}-\hbar k_{\rm i})\right]^2.
\end{eqnarray}
This provides the boundary conditions 
\begin{eqnarray}
M\dot{R}(t_{\rm f})=\hbar k_{\rm f}=\sqrt{2mE'}, \ \ 
M\dot{R}(t_{\rm i})=\hbar k_{\rm i}=\sqrt{2mE}
\label{bcon}
\end{eqnarray}
on functional integrals in Eq.~(\ref{2B.14}).
With a proviso that these conditions are met, we obtain
\begin{eqnarray}
\Gamma(E,E')
&=&\!\lim_{t_{\rm f,i}\rightarrow\pm\infty}\frac{1}{t_{\rm f}-t_{\rm i}}
\!\int^\infty_{-\infty}\!\!\frac{dt}{2\pi\hbar}e^{\frac{i}{\hbar}(E-E')t}
\int \!\!\!\! \int^{X(t_{\rm f})=X_{\rm f}}_{X(t_{\rm i})=X_{\rm i}} \!\!\!
{\cal D}X \!\!
\int \!\!\!\! \int^{Y(t_{\rm f})=Y_{\rm f}}_{Y(t_{\rm i})=Y_{\rm i}}\!\!\!
{\cal D}Y  
\exp\!\left[\frac{i}{\hbar}S_{\rm eff}(t)\right]\!.
\label{2B.15}
\end{eqnarray}
This result together with Eq.~(\ref{2B.10}) or (\ref{2B.11}) gives the desired {\it exact} expression of the transmission spectrum of a particle interacting with a bath of harmonic oscillators.

The integral of $\Gamma(E,E')$ over $E'$ gives the {\it total} tunneling rate $\Gamma_{\rm total}(E)$ for a particle with initial-state energy $E$. From Eq.~(\ref{2B.15}) we obtain
\begin{eqnarray}
\Gamma_{\rm total}(E)&=&\!\int^\infty_{-\infty}\!\Gamma(E,E')dE' 
=\!
\lim_{t_{\rm f,i}\rightarrow\pm\infty}\frac{1}{t_{\rm f}-t_{\rm i}}
\int \!\!\!\!\! \int^{X(t_{\rm f})=X_{\rm f}}_{X(t_{\rm i})=X_{\rm i}} \!\!\! {\cal D}X \!\!
\int \!\!\!\!\! \int^{Y(t_{\rm f})=Y_{\rm f}}_{Y(t_{\rm i})=Y_{\rm i}} \!\!\! {\cal D}Y 
\exp\!\left[\frac{i}{\hbar}S_{\rm eff}(t=0)\right]\!.
\label{2B.16}
\end{eqnarray}
Thus the total tunneling rate is given by the value of $S_{\rm eff}(t)$ at $t=0$.
This action should be related to the imaginary-time effective action derived by Caldeira and Leggett using the instanton method because the latter also gives the total tunneling rate for the same system. We will show this in Sec.~\ref{sec:Instanton}.

The physics contained in the effective action~(\ref{2B.11}) is revealed if we rewrite it as $S_{\rm eff}(t)=S_{\rm eff}(t=0)+[S_{\rm eff}(t)-S_{\rm eff}(t=0)]$:
\begin{eqnarray}
S_{\rm eff}(t)\!&=&\!S_{\rm eff}(t=0)
\!+\!i\!\!\int^{t_{\rm f}}_{t_{\rm i}}\!\!dt_1\int^{t_{\rm f}}_{t_{\rm i}}\!\!dt_2\!\int^\infty_0\!\!\frac{d\omega}{\pi}\frac{J(\omega)}{\omega^2}
\dot{g}[X(t_1)]\dot{g}[Y(t_2)]
\nonumber \\
& & \!\times
\left\{\left[n_{\rm B}(\omega)+1\right](1\!-\!e^{-i\omega t})e^{i\omega(t_1-t_2)}
 \!+\!n_{\rm B}(\omega) (1\!-\!e^{i\omega t} )e^{-i\omega(t_1-t_2)}
\right\}\!,
\label{2B.17}
\end{eqnarray}
where $n_{\rm B}(\omega)\equiv(e^{\beta\hbar\omega}-1)^{-1}$ is the boson occupation number.
As shown in Eq.~(\ref{2B.16}) the first term $S_{\rm eff}(t=0)$ on the right-hand side of Eq.~(\ref{2B.17}) describes the total tunneling rate.
The second term comprises two ingredients in the curly brackets: the first of them describes emission of the energy quanta and the second describes absorption of energy quanta.
At zero temperature, the second one vanishes because the bath has no energy to give off, whereas the first one survives because of the presence of spontaneous emission.
The presence of spontaneous emission is a unique feature of the interaction with dynamical fields and makes a distinction from the interaction with classical fields as will be illustrated in Sec.~\ref{sec:bltime}.

\section{Stationary-Phase Approximation}
\label{sec:Stationary}

\subsection{Equations of motion}

We evaluate the functional integrals in (\ref{2B.15}) by means of stationary-phase approximation in which the functional integrals are replaced by ordinary integrals along the stationary-phase trajectories.
Assuming that the particle interacts with the bath only under the barrier, we choose
\begin{eqnarray}
f_\alpha(X)
=\cases{0                                 &if $X<0$;          \cr
\frac{C_\alpha}{m_\alpha\omega_\alpha^2}X &if $0\leq X\leq d$;\cr
\frac{C_\alpha}{m_\alpha\omega_\alpha^2}d &if $X>d$, \cr}
\label{condition}
\end{eqnarray}
where $0$ and $d$ are the left and right turning points of the barrier.
Since the particle must be on one side (or the other side) of the barrier in the initial state (or the final state), $f_\alpha(X)$ given in Eq.~(\ref{condition}) indeed satisfies the condition (\ref{assumption2}).
The first condition~(\ref{assumption1}) is also satisfied by our choice of the potential described later.

The fact that the Hamiltonian~(\ref{2B.1}) does not depend explicitly on time means that the system is invariant under translation of time.
The quantum fluctuations around the stationary-phase trajectories therefore include the zero-frequency mode which produces the prefactor proportional to $t_{\rm f}-t_{\rm i}$~\cite{Langer,Coleman}.
Contributions of the remaining quantum fluctuations produces an extra prefactor $N$ which, in general, depends on energy, $T_X$, and $T_Y$, etc.
However, we assume that in the semiclassical limit its dependence is weaker than that of the tunneling exponent, and we will neglect its dependence in the following discussions.
We thus find from Eq.~(\ref{2B.15}) that
\begin{eqnarray}
\Gamma(E,E')
=N\int^\infty_{-\infty}\!\!\frac{dt}{2\pi\hbar}e^{\frac{i}{\hbar}(E-E')t}
e^{\frac{i}{\hbar}S(t)}\!,
\label{3A.1}
\end{eqnarray}
where $S(t)$ is obtained from $S_{\rm eff}(t)$ by
replacing the upper limits of integration by $T_X$ (or $T_Y$) for trajectory $X$ (or $Y$) and the lower limits of integration by $0$.
The $T_X$ and $T_Y$ will later be determined so as to meet conditions
$X(T_X)=Y(T_Y)=d$, provided that $X(0)=Y(0)=0$,
and may be regarded as (complex) traversal times in the sense that they are the ^^ ^^ times" it takes the particle to traverse the barrier along the stationary-phase trajectories $X$ and $Y$.
At zero temperature $S(t)$ becomes
\begin{eqnarray}
S(t)\!&=&\!\!\int^{T}_0\!\!\!dt_1\!
\left[\frac{M}{2}\dot{X}^2\!\!-\!\frac{M}{2}\dot{Y}^2\!\!-\!V(X)\!+\!V(Y)\!-\!\frac{\mu}{2}X^2\!+\!\frac{\mu}{2}Y^2\right]
+ i\!\int^T_0\!\!\!dt_1\!\int^T_0\!\!\!dt_2\!\!\int^\infty_0\!
\frac{d\omega}{2\pi}J(\omega)
\nonumber \\
& & \times
\left[
e^{-i\omega|t_1-t_2|}X(t_1)X(t_2)\!+\!e^{i\omega|t_1-t_2|}Y(t_1)Y(t_2)
-2e^{i\omega[(t_1-T_X)-(t_2-T_Y)]}X(t_1)Y(t_2)
\right]\!\!
\nonumber \\
& & + \ id^2\!\!\!\int^\infty_0\!\!\frac{d\omega}{\pi}
\frac{J(\omega)}{\omega^2}(1\!-\!e^{-i\omega t})
\left[1\!-\!i\omega\!\int^T_0\!\!\!dt_1e^{i\omega(t_1-T)}\frac{X(t_1)}{d}\right]\!\!
\left[1\!+\!i\omega\!\int^T_0\!\!\!dt_2e^{i\omega(t_2-T)}\frac{Y(t_2)}{d}\right]\!,
\label{3A.2}
\end{eqnarray}
where it is understood that $T$ is equal to $T_X$ (or $T_Y$) if the trajectory referred to is $X$ (or $Y$), and
\begin{eqnarray}
\mu\equiv\sum_\alpha\frac{C_\alpha^2}{m_\alpha\omega_\alpha^2}=\frac{2}{\pi}
\int^\infty_0\!\frac{J(\omega)}{\omega}d\omega.
\label{3A.3}
\end{eqnarray}

In Eq.~(\ref{3A.2}) we have replaced the factor $e^{i\omega(t_1-t_2)}$ in front of $X(t_1)Y(t_2)$ by $e^{i\omega[(t_1-T_X)-(t_2-T_Y)]}$.
This apparently {\it ad hoc} replacement must be made in order to get physically reasonable results in the limit of $\omega T_0\rightarrow\infty$ (see Sec.~\ref{sec:Single}).
In fact, in the limit of large $\omega T_0$, the only effect of the interaction between the system and the bath is the potential renormalization that should be canceled by the counter term.
Thus there should be no effect of the interaction on the tunneling rate,
which will be confirmed by the above replacement.
If we could perform the functional integrals exactly, the original factor $e^{i\omega(t_1-t_2)}$ should, of course, give the correct answer.
In resorting to the saddle-point approximation in real time, however, we would have unphysical results with the original factor.
It will turn out that the above replacement also reproduces the same total tunneling rate as is obtained by the instanton method for arbitrary $\omega T_0$ (see Sec.~\ref{sec:Instanton}) as well as the same transmission spectrum that is obtained by the transfer-Hamiltonian method~\cite{Devoret} in the limit of $\omega T_0\rightarrow 0$ (see Sec.~\ref{sec:Small}).

The stationary-phase trajectories for $X$ and $Y$ are determined so as to make $S(t)$ extremal. 
From conditions $\delta S(t)/\delta X(t_1)=0$ and $\delta S(t)/\delta Y(t_1)=0$, we obtain the equations of motion for $X$ and $Y$:
\begin{eqnarray}
M\ddot{X}_1\!&=&\!-\frac{\partial V}{\partial X(t_1)}\!-\!\mu X(t_1)
\!+\!i\!\int^\infty_0\!\!\frac{d\omega}{\pi}J(\omega)\!
\left[\int^{T_X}_0\!\!\!\!\!\!dt_2e^{-i\omega|t_1-t_2|}X(t_2)\!
 -\!\!\int^{T_Y}_0\!\!\!\!\!\!dt_2e^{i\omega[(t_1-T_X)-(t_2-T_Y)-t]}Y(t_2)
\!\right]
\nonumber \\
& & + d\!\int^\infty_0\!\!\frac{d\omega}{\pi}
\frac{J(\omega)}{\omega}(1-e^{-i\omega t})e^{i\omega(t_1-T_X)};
\nonumber \\
M\ddot{Y}_1\!&=&\!-\frac{\partial V}{\partial Y(t_1)}\!-\!\mu Y(t_1)
\!+\!i\!\int^\infty_0\!\!\frac{d\omega}{\pi}J(\omega)\!
\left[\int^{T_Y}_0\!\!\!\!\!\!dt_2e^{i\omega|t_1-t_2|}Y(t_2)\!
 -\!\!\int^{T_X}_0\!\!\!\!\!\!dt_2e^{-i\omega[(t_1-T_Y)-(t_2-T_X)+t]}X(t_2)\right]
\nonumber \\
& & - d\!\int^\infty_0\!\!\frac{d\omega}{\pi}
\frac{J(\omega)}{\omega}(1-e^{-i\omega t})e^{-i\omega(t_1-T_Y)}.
\label{3A.4}
\end{eqnarray}
To uniquely determine the solutions to Eqs.~(\ref{3A.4}), we must specify initial velocities $\dot{X}(0)$ and $\dot{Y}(0)$. 
Multiplying the first equation in (\ref{3A.4}) by $\dot{X}(t_1)$ and
integrating it from $t_1=0$ to $t_1=t'$, we obtain obtain the law of energy conservation:
\begin{eqnarray}
& & \frac{M}{2}\dot{X}^2(0)+V(0)=
\frac{M}{2}\dot{X}^2(t')
+U_{\rm eff}[X(t')],
\label{3A.6}
\end{eqnarray}
where 
\begin{eqnarray}
U_{\rm eff}[X(t')]\!&=&\!V[X(t')]+\frac{\mu}{2}X^2(t')
-i\int^{t'}_0dt_1
\!\int^\infty_0\!\!\frac{d\omega}{\pi}J(\omega)
\nonumber \\
& & \times
\left[\int^{T_X}_0\!\!\!\!\!dt_2e^{-i\omega|t_1-t_2|}X(t_2)\!
-\!\!\int^{T_Y}_0\!\!\!\!\!dt_2e^{i\omega[(t_1-T_X)-(t_2-T_Y)-t]}Y(t_2)\right]
\nonumber \\
& &
+ d\!\int^{t'}_0\!\!dt_1\dot{X}(t_1)
\!\int^\infty_0\!\!\frac{d\omega}{\pi}
\frac{J(\omega)}{\omega}(1-e^{-i\omega t})e^{i\omega(t_1-T_X)}
\label{3A.7}
\end{eqnarray}
is an effective potential which is a local function of time because the
stationary-phase trajectories $X$ and $Y$ are uniquely determined as
functions of time.
To determine initial velocity $\dot{X}(0)$, we must specify the value of
the left-hand side of Eq.~(\ref{3A.6}), 
i.e., the initial energy of a tunneling particle. 
To be consistent with the WKB result, we should identify it
with the incident energy $E$ of the particle.
We thus obtain $\dot{X}(0)=\pm i\sqrt{2[V(0)-E]/M}$.
To ensure that the wave function does not grow but decay inside the barrier,
we take the plus sign.
Initial velocity $\dot{Y}(0)$ for trajectory $Y$ can be determined along a similar line, giving
\begin{eqnarray}
\dot{X}(0)=-\dot{Y}(0)=i\sqrt{\frac{2[V(0)-E]}{M}} \equiv i\frac{d}{T_0},
\label{3A.8}
\end{eqnarray}
where the last equation defines $T_0$.

\subsection{Analytic solutions}

The integro-differential equations (\ref{3A.4}) can be solved analytically
as long as they are linear in $X$ and $Y$.
This means that analytic solutions can be obtained if $V(X)$ has the form
\begin{eqnarray}
V(X)=\cases{0                                     &if $X<0$;          \cr
U_0+U_1\frac{X}{d}+U_2\left(\frac{X}{d}\right)^2  &if $0\leq X\leq d$;\cr
            0                                     &if $X>d$. \cr}
\label{3B.1}
\end{eqnarray}
In what follows, we will discuss the case of $U_2=0$ to avoid 
making solutions unnecessarily complicated. (The case of $U_2\neq 0$ can be
treated along a similar line to what is described below.)
The type of the potential with $U_2=0$ appears, for instance, 
when electrons tunnel through a square potential of height
$U_0$ which is tilted by voltage bias $V$, i.e., $U_1=-eV$.
Integrating Eqs.~(\ref{3A.4}) twice with respect to time yields
\begin{eqnarray}
\frac{X(t_1)}{d}\!&=&\!i\frac{t_1}{T_0}+\!\frac{u_1}{2}t_1^2
-\!\frac{2}{M}\!\!\int^{t_1}_0\!\!\!\!dt'
\!\!\int^\infty_0\!\!\frac{d\omega}{\pi}
\frac{J(\omega)}{\omega^2}\sin\omega(t_1\!-\!t')
\nonumber \\
& & 
-\frac{1}{M}\int^\infty_0\frac{d\omega}{\pi}\frac{J(\omega)}{\omega^3}
(e^{i\omega t_1}-i\omega t_1-1)a_\omega(t),
\nonumber \\
\frac{Y(t_1)}{d}\!&=&\!-i\frac{t_1}{T_0}+\!\frac{u_1}{2}t_1^2
-\!\frac{2}{M}\!\!\int^{t_1}_0\!\!\!\!dt'
\!\!\int^\infty_0\!\!\frac{d\omega}{\pi}
\frac{J(\omega)}{\omega^2}\sin\omega(t_1\!-\!t')
\nonumber \\
& & 
-\frac{1}{M}\int^\infty_0\frac{d\omega}{\pi}\frac{J(\omega)}{\omega^3}
(e^{-i\omega t_1}+i\omega t_1-1)b_\omega(t),
\label{3B.2}
\end{eqnarray}
where $u_1\equiv U_1/Md^2$ and
\begin{eqnarray}
& & 
a_\omega(t)=i\omega\!\!\int^{T_X}_0\!\!\!\!\!dt'e^{-i\omega t'}\frac{X(t')}{d}
-i\omega\!\!\int^{T_Y}_0\!\!\!\!\!dt'e^{-i\omega(t'+t+T_X-T_Y)}\frac{Y(t')}{d}\!+\!(1\!-\!e^{-i\omega t})e^{-i\omega T_X},
\nonumber \\
& & 
b_\omega(t)=-i\omega\!\!\int^{T_X}_0\!\!\!\!\!dt'e^{i\omega t'}\frac{Y(t')}{d}
+i\omega\!\!\int^{T_X}_0\!\!\!\!\!dt'e^{i\omega(t'-t-T_X+T_Y)}\frac{X(t')}{d}\!+\!(1\!-\!e^{-i\omega t})e^{i\omega T_Y}.
\label{3B.3}
\end{eqnarray}
From Eqs.~(\ref{3B.2}) and (\ref{3B.3}) we find that the solutions have the following general properties:
\begin{eqnarray}
X(t_1)&=&Y(-t_1), \nonumber \\
a_\omega(t)&=&b_\omega(t), \nonumber \\
T_X&=&-T_Y.
\label{3B.4}
\end{eqnarray}
These properties can be used to reduce the coupled integro-differential equations to a single one, i.e., the first equation of Eqs.~(\ref{3B.2}),  which is solved by the Laplace transformation.
The Laplace transform $\tilde{X}(s)$ of $X(t')$ is given by
\begin{eqnarray}
\frac{\tilde{X}(s)}{d}
\!&=&\!
\frac{\frac{i}{T_0}s\!+\!u_1\!+\frac{1}{M}
\int^\infty_0\!\!\frac{d\omega}{\pi}\frac{J(\omega)}{\omega}\frac{s}{s-i\omega}
a_\omega(t)}
{s^3\left[1+\frac{2}{M}\int^\infty_0\frac{d\omega}{\pi}\frac{J(\omega)}{\omega}
\frac{1}{s^2+\omega^2}\right]}, 
\label{3B.5}
\end{eqnarray}
where 
\begin{eqnarray}
a_\omega(t)&=&i\omega\!\!\int^T_0\!\!\!\!dt'
\left[
e^{-i\omega t'}+e^{-i\omega(t+t'-2T)}
\right]\frac{X(t')}{d}
+(1-e^{-i\omega t})e^{-i\omega T}.
\label{3B.6}
\end{eqnarray}
Inverse Laplace transformation of Eq.~(\ref{3B.5}) gives solution $X(t')$
which includes $a_\omega(t)$ and $T$ as undetermined.
Here $a_\omega(t)$ is determined by substitution of the obtained solution into Eq.~(\ref{3B.6}) and $T$ is determined from condition $X(T)=d$.
Equations~(\ref{3B.4})-(\ref{3B.6}) therefore give the exact solutions for the stationary-phase trajectories for an arbitrary spectral density function of the bath.

\subsection{Tunneling exponent}

Integrating by parts the kinetic energy terms in
Eq.~(\ref{3A.2}) and eliminating the resulting second time derivatives of $X$ and $Y$ using Eqs.~(\ref{3A.4}), we find that
\begin{eqnarray}
S(t)\!&=&\!\frac{Md}{2}\!\left[\dot{X}(T_X)\!-\!\dot{Y}(T_Y)\right]
\!+\!\frac{1}{2}\!\int^{T_X}_0\!\!\!dt_1\!\!
\left[X\frac{\partial V}{\partial X}\!-\!2V(X)\right]
-\frac{1}{2}\!\int^{T_Y}_0\!\!\!dt_1\!\!
\left[Y\frac{\partial V}{\partial Y}\!-\!2V(Y)\right]
\nonumber \\
& & 
\!+id\!\int^T_0\!\frac{d\omega}{\pi}\frac{J(\omega)}{\omega^2}
(1-e^{-i\omega t})
\left\{\!1\!-\!\frac{i\omega}{2}\!\!\int^T_0\!\!\!dt_1\!\!
\left[e^{i\omega(t_1-T)}\frac{X(t_1)}{d}\!-\!e^{-i\omega(t_1-T)}\frac{Y(t_1)}{d}\right]\!\right\}\!.
\label{3C.1}
\end{eqnarray}
Relations (\ref{3B.4}) can be used to simplify this as 
\begin{eqnarray}
& & \frac{S(t)}{S_0}\!=\!\frac{\dot{X}(T)T_0}{2d}
-\frac{T}{2T_0}
+\frac{U_1}{S_0}\!\int^T_0\!\!\!dt_1\frac{X(t_1)}{d}
+\frac{iT_0}{2M}\!\int^T_0\!\!\frac{d\omega}{\pi}\frac{J(\omega)}{\omega^2}
\left[1-i\omega\int^T_0\!\!\!dt_1e^{i\omega(t_1-T)}\frac{X(t_1)}{d}
\right],
\label{3C.2}
\end{eqnarray}
where
\begin{eqnarray}
S_0=4(U_0-E)T_0
\label{3C.3}
\end{eqnarray}
is the tunneling exponent in the absence of the interaction.

Equations~(\ref{3B.5}), (\ref{3B.6}) and (\ref{3C.2}) constitutes the complete solution to the problem.

\subsection{Relation to the instanton method}
\label{sec:Instanton}

The effective action $S_{\rm eff}(t)$ can be related to the effective action obtained by the instanton method.
The instanton method imposes the boundary conditions $X(0)=X(2T)=0$. To satisfy them, we replace time variables $t_i-T$ ($i=1,2$) by $-i(\tau_i-T)$ for trajectory $X(t_i)$ and by $i(\tau_i-T)$ for trajectory $Y(t_i)$.
The resulting effective action, which we denote as $S^{\rm B}(t)$, reads
\begin{eqnarray}
& & S^{\rm B}(t)=\int^{2T}_0\!\!\!\!d\tau\left[\frac{M}{2}\dot{X}^2+V(X)+\mu X^2(\tau)\right]
-\int^{2T}_0\!\!\!\!d\tau_1\int^{2T}_0\!\!\!\!d\tau_2 \ \alpha(\tau_1-\tau_2)X(\tau_1)X(\tau_2)
\nonumber \\
& & +d^2\int^\infty_0\!\!\frac{d\omega}{2\pi}\left\{
n_{\rm B}(-\omega)(1-e^{-i\omega t})
\left[1-\omega\int^T_0\tau_1e^{\omega(\tau_1-T)}\frac{X(\tau_1)}{d}\right]^2+(\omega\longrightarrow -\omega)
\right\},
\label{imaginary}
\end{eqnarray}
where $\alpha(\tau)=\alpha^>(t=-i|\tau|)$ and the second term in the curly braces can be obtained by replacing $\omega$ in the preceding term by $-\omega$.
As shown in Eq.~(\ref{2B.16}), the effective action that gives the total tunneling rate is given by setting $t=0$ in $S^{\rm B}(t=0)$.
Then the last term in Eq.~(\ref{imaginary}), which is responsible for decomposing the total escape rate into transfer energies, vanishes and the remaining action takes the form similar to the one obtained by Caldeira and Leggett:~\cite{CL}
\begin{eqnarray}
& & S^{\rm CL}=\int^{\beta\hbar}_0\!\!\!\!d\tau\left[\frac{M}{2}\dot{X}^2+V(X)\right]
+\int^{\beta\hbar}_0\!\!\!\!d\tau_1\int^{\beta\hbar}_0\!\!\!\!d\tau_2 \ \alpha(\tau_1-\tau_2)\left[X(\tau_1)-X(\tau_2)\right]^2.
\label{CL}
\end{eqnarray}
Different upper bounds between $S^{\rm B}$ and $S^{\rm CL}$ result from different boundary conditions.
In the former case, the boundary condition $X(0)=X(2T)$ is imposed, while in the latter case the thermodynamic boundary condition $X(0)=X(\hbar\beta)$ is imposed.  

In the following two sections we will use these results to compute the transmission spectra for a bath with single frequency and for a bath with Ohmic spectrum.
\section{Transmission spectrum for a bath with single frequency}
\label{sec:Single}

The transmission spectrum for a bath with single frequency has been studied for the case with and without the counter term in Refs.~\cite{UA2} and \cite{UA1} using an imaginary-time approach.
This section has two aims: One is to confirm that the results obtained by the imaginary-time approach~\cite{UA2} are reproduced by the real-time approach presented in the preceding sections.
The other is to derive a new analytic expression of the tunneling exponent
which covers from sudden tunneling ($\omega T_0\ll 1$) to adiabatic one 
($\omega T_0\gg 1$).
These two regimes were treated separately in Refs.~\cite{UA2}.

\subsection{Stationary-phase solution}

The spectral density function of a bath with single frequency $\omega$ is given by
\begin{eqnarray}
J(\omega')=\pi M\gamma\omega^3\delta(\omega'-\omega),
\label{4A.1}
\end{eqnarray}
where $\gamma$ is a dimensionless coupling constant.
Substituting Eq.~(\ref{4A.1}) into Eq.~(\ref{3B.5}) and inversely
Laplace-transforming it gives
\begin{eqnarray}
\frac{X(t_1)}{d}\!&=&\!\frac{1}{p^2}\!\left[\frac{i}{T_0}\!\left(\!t_1+2\gamma\frac{\sin p\omega t_1}{p\omega}\!\right)+\frac{u_1}{2}\!
\left(t_1^2+4\gamma\frac{\!1\!-\!\cos p\omega t_1\!}{(p\omega)^2}\right)
\right. \nonumber \\
& &\left.+ \gamma a_\omega(t)\!\left(1-\cos p\omega t_1+i\omega t_1-i \frac{\sin p\omega t_1}{p}\right)\right],
\label{4A.2}
\end{eqnarray}
where $p\equiv\sqrt{1+2\gamma}$.
Substituting Eq.~(\ref{4A.2}) into Eq.~(\ref{3B.6}) and solving for $a_\omega(t)$, we find that
\begin{eqnarray}
& & a_\omega(t)= -\frac{1}{\omega T_0}
\frac{2(1\!-\!\cos p\omega T)\!+\!\frac{4iU_1}{p\omega S_0}\!
\left(\sin p\omega T\!-\!p\omega T\right)\!-\!F(\omega)(1\!-\!e^{-i\omega t})}
{\cos p\omega T\!+\!ip\sin p\omega T\!+\!2\gamma\!
-\! \frac{i\gamma}{p} \left(\sin p\omega T\!-\!p\omega T\right)\!(1\!-\! e^{-i\omega t})},
\nonumber \\
\label{4A.3}
\end{eqnarray}
where
\begin{eqnarray}
F(\omega)&\!=\!& 1\!-\!\cos p\omega T\!+\!i\frac{\sin p\omega T}{p}\!-\!
i\omega T + p^2\omega T
\nonumber \\
& &
+\frac{2U_1}{p^2\omega S_0}
\left[
1\!-\!\frac{(p\omega T)^2}{2}-\!\cos p\omega T\!+
ip\sin p\omega T\!-\!ip^2\omega T \right].
\label{4A.4}
\end{eqnarray}

\subsection{Small-$\gamma$ expansion}
\label{sec:Small}

When $\gamma \ll 1$, it is sufficient to keep
terms up to the first order in $\gamma$. Neglecting terms of the order of
$\gamma^2$ and those of the order of $\gamma U_1/U_0$, we find that
\begin{eqnarray}
\frac{X(t_1)}{d}\!&=&\!\frac{i}{T_0}\!\left[
(1\!-\!2\gamma)t_1\!+\!2\gamma\frac{\sin \omega t_1}{\omega}\right]\!
+\!\frac{u_1}{2}t_1^2
+\gamma a_\omega(t)\left(1\!-\!\cos \!\omega t_1\!+\!i\omega t_1\!-\!i\sin \!\omega t_1\right)\!.
\label{4B.1}
\end{eqnarray}
Since the last term in Eq.~(\ref{4B.1}) is already multiplied by $\gamma$, it is
sufficient to expand Eq.~(\ref{4A.3}) up to the zeroth orde in $\gamma$:
\begin{eqnarray}
a_\omega(t)=\frac{e^{-\omega T_0}}{\omega T_0}\!
\left[ 2(\cosh\!\omega T_0-1)+(1-e^{-\omega T_0})(1-e^{-i\omega t})\right]\!.
\label{4B.2}
\end{eqnarray}
To evaluate the tunneling exponent in Eq.~(\ref{3C.2}), we need to find $T$.
Since $\gamma$ is small, it is reasonable to assume that $T/T_0$ can
be expanded in powers of $\gamma$:
\begin{eqnarray}
\frac{T}{T_0}=-i(1+x),
\label{4B.3}
\end{eqnarray}
where $x$ is of the order of $\gamma$ or $U_1/U_0$.
From condition $X(T)=d$ and Eqs.~(\ref{4B.1})-(\ref{4B.3}),
we find that
\begin{eqnarray}
\frac{T}{T_0}\!&=&\!-i\left\{
1\!+\!\frac{U_1T_0}{S_0}
\!+\!\frac{\gamma}{\omega T_0}
\!\left[2(e^{-\omega T_0}\!\!-\!1\!+\!\omega T_0)\!-\!(1\!-\!e^{-\omega T_0})^2
(1\!+\!\omega T_0)
\right.\right.
\nonumber \\
& & 
\left.\left.
+(1\!-\!e^{-\omega T_0})(1\!-\!e^{-\omega T_0}\!\!-\!\omega T_0 e^{-\omega T_0})(1\!-\!e^{-i\omega t})\right]\!\right\}\!.
\label{4B.4}
\end{eqnarray}
Substituting Eqs.~(\ref{4B.1})-(\ref{4B.4}) into Eq.~(\ref{3C.2}) yields
\begin{eqnarray}
\frac{S(t)}{S_0}\!&=&\!i\left[1\!-\!\frac{U_1T_0}{S_0}
\!+\!\gamma D(\omega T_0)
\!+\!\gamma W(\omega T_0)(1\!-\!e^{-i\omega t})
\right]\!,
\label{4B.5}
\end{eqnarray}
where
\begin{eqnarray}
D(x)\!&=&\!1\!-\!\frac{1\!-\!e^{-x}}{2x}(3\!-\!e^{-x}) 
=\cases{\frac{x^2}{3}\!-\!\frac{x^3}{4} &if $x\ll 1$; \cr
1\!-\!\frac{3}{2x}                      &if $x\gg 1$, \cr}
\label{4B.6}
\end{eqnarray}
and
\begin{eqnarray}
W(x)=\frac{(1\!-\!e^{-x})^2}{2x}
=\cases{\frac{x}{2}\!-\!\frac{x^2}{2}  &if $x\ll 1$; \cr
\frac{1}{2x}                           &if $x\gg 1$. \cr}
\label{4B.7}
\end{eqnarray}
Substituting Eq.~(\ref{4B.5}) into Eq.~(\ref{3A.1}) yields the desired transmission spectrum
\begin{eqnarray}
\Gamma(E,E')\!&=&\!N
\exp\left\{-\frac{S_0}{\hbar}\left[1-\frac{U_1T_0}{S_0}+\gamma D(\omega T_0)\right]\right\}
\nonumber \\
& & \times e^{-\gamma\frac{S_0}{\hbar}W(\omega T_0)}
\sum^\infty_{n=0}\frac{[\gamma\frac{S_0}{\hbar}W(\omega T_0)]^n}{n!}
\delta(E-E'-n\hbar\omega).
\label{4B.8}
\end{eqnarray}
Here the first exponent gives the total tunneling rate while the remaining term shows how the total tunneling rate is distributed over final states having different energies.
Each channel $n$ corresponds to a process in which $n$ quanta of the bath are excited by the tunneling process.
The first exponent on the right-hand side of Eq.~(\ref{4B.8}) comprises three ingredients in the square bracket:
1 gives the WKB exponent in the absence of the interaction,
$-U_1T_0/S_0$ gives the contribution from the bias, and
$\gamma D(\omega T_0)$ describes the effect of the interaction.
The positive sign of $D(\omega T_0)$ implies that the total tunneling rate is suppressed by the interaction.

Since $D(\omega T_0)$ becomes unity in the limit of $\omega T_0\rightarrow\infty$, the effect of the interaction does not appear to vanish in this limit, although  the original Hamiltonian~(\ref{2B.1}) appears to require that it should vanish because of the presence of the counter term
\begin{eqnarray}
H_{\rm counter}=\sum_\alpha\frac{C_\alpha^2}{2m_\alpha\omega_\alpha^2}X^2
=\frac{X^2}{\pi}\int^\infty_0\frac{J(\omega')}{\omega}d\omega.
\label{4B.9}
\end{eqnarray}
Since the counter term gives the maximum possible interaction energy, the ratio of it to the bare potential $U_0$ may be regarded as the ^^ ^^ physical" coupling constant $g$ which is to be held constant. 
Noting that the counter term scales as $(X/d)^2$, we are led to define $g$ as
\begin{eqnarray}
g\equiv\sqrt{\frac{H_{\rm counter}}{U_0\left(\frac{X}{d}\right)^2}}.
\label{4B.10}
\end{eqnarray}
For the spectral density function in Eq.~(\ref{4A.1}) we have
\begin{eqnarray}
g=\sqrt{2\gamma(\omega T_0)^2} \ \ \ {\rm or} \ \ \ \gamma=\frac{g^2}{2(\omega T_0)^2}.
\label{4B.11}
\end{eqnarray}
If we take the limit $\omega T_0\rightarrow\infty$ with $g$ held constant,
the effect of the interaction vanishes as $\sim(1/\omega T_0)^2$ in the limit of $\omega T_0\rightarrow\infty$, whereas it remains finite if $g$ is increased in proportion to $\omega T_0$ as seen above.
It is interesting, however, to observe that the single parameter that characterizes the tunneling exponent is $\gamma$ and not $g$.
In the physics of small-capacitance tunnel junctions, $g^2$ gives the ratio of the elementary charging energy to the bare barrier height~\cite{UA1}.
Figure~1 shows the normalized tunneling exponent $S/S_0$ as a function of $\omega T_0$ for several values of $g$, where $S=S_{\rm eff}(t=0)$.
The total tunneling rate decreases with increasing coupling constant $g$ or decreasing $\omega T_0$.

\subsection{Tunneling through a fluctuating barrier: classical vs. quantal modulations}
\label{sec:bltime}

Quantum tunneling through a fluctuating barrier has seen a remarkable resurgence of interest since B\"{u}ttiker and Landauer~\cite{BL} reconsidered the problem of tunneling through a barrier that is time-modulated in a prescribed manner.
For an opaque barrier,
they found that the upper and lower sideband intensities,
$I_+$ and $I_-$, of transmitted particles having incident energy $E$ satisfy the following relation:
\begin{eqnarray}
\frac{I_+-I_-}{I_++I_-}=\tanh\omega T_0,
\label{4C.1}
\end{eqnarray}
where $\omega$ is the frequency of the modulation and 
$T_0$ is given by
\begin{eqnarray}
T_0=\int^d_0 \!\!\frac{dx}{\sqrt{\frac{2}{M}[V(x)-E]}}.
\label{4C.2}
\end{eqnarray}
For $\omega T_0\ll 1$ a tunneling particle will see an instantaneous potential, while for $\omega T_0\gg 1$ it will see a time-averaged one.
One can therefore expect a crossover between the two regimes around $\omega T_0=1$.
In fact, as the frequency of the modulation increases, the ratio (\ref{4C.1}) rapidly increases from 0 to 1 around $\omega T_0=1$, which suggests that a time scale for the traversal time for tunneling is given by the inverse characteristic frequency at which the crossover occurs.

The crossover argument provided by B\"{u}ttiker and Landauer is relevant to many problems of physical interest such as those of the dynamic image potential~\cite{Jonson,Persson,Sebastian,Gueret}, those of tunneling through a Josephson junction~\cite{Esteve}, and more recently those of tunneling through smal-capacitance normal tunnel junctions that could operate at room temperature~\cite{UA1,UA3}.
In these problems, however, barrier modulations are caused not by classical sources but by dynamical degrees of freedom contributed mostly from their zero-point fluctuations.
In the problem of dynamic image potential, for example, the barrier modulation is caused by the zero-point fluctuations of surface plasmon modes~\cite{Sebastian}.
In the problems of tunneling through small-capacitance normal and superconducting junctions, barrier modulations are caused by quantum fluctuations of the electromagnetic (EM) environment surrounding the junction~\cite{Nazarov,Devoret,Girvin}.
We use the results obtained thus far to study the case in which the barrier modulation is caused by the quantum-mechanical and thermal fluctuations of the environmental degrees of freedom.
When the coupling of the particle to the bath is not strong, we find from Eqs.~(\ref{2B.17}) and (\ref{4A.1}) that
\begin{eqnarray}
& & \Gamma(E,E')\propto 
\sum^\infty_{n=-\infty} \!\!I_n\!\!\left[\frac{h(\omega)}{\sinh\!\frac{\beta\hbar\omega}{2}} \right]e^{n\omega(T_0-\beta\hbar/2)} 
\delta(E'-E-n\hbar\omega),
\label{4C.3}
\end{eqnarray}
where $I_n$ is the modified Bessel function, and
$h(\omega)=(\gamma S_0/\hbar\omega T_0)\sinh^2\!(\omega T_0/2)$.
For the n-th upper and lower sidebands $I_{\pm n}$ we find that
\begin{eqnarray}
\frac{I_{+n}-I_{-n}}{I_{+n}+I_{-n}}=\tanh n\omega \left(T_0-\frac{\beta\hbar}{2}\right).
\label{4C.4}
\end{eqnarray}
In the high-temperature limit where $\beta\rightarrow 0$, Eq.~(\ref{4C.4}) reduces to the result obtained for classical modulation, esp., a specific result for $n=1$ coincides with that obtained in Ref.~\cite{BL}.
In the low-temperature limit where $\beta\rightarrow\infty$, however, Eq.~(\ref{4C.4}) approaches -1.
Physically this is because the particle cannot absorb energy from the bath at zero temperature.
Thus the crossover argument for the traversal time for tunneling does not hold true when the barrier modulation is caused mainly by quantum-mechanical zero-point fluctuations.
A special case of $n=1$ is illustrated in Fig.~2 for various values of the ratio of the energy quantum $\hbar\omega$ of the bath to the thermal energy $\beta^{-1}$.

\section{Transmission spectrum for a bath with Ohmic spectrum}
\label{sec:Ohmic}

\subsection{Stationary-phase solution}

The spectral density function for a bath with Ohmic spectrum is given by
\begin{eqnarray}
J(\omega)=M\eta\omega.
\label{5A.1}
\end{eqnarray}
Substituting Eq.~(\ref{5A.1}) into Eq.~(\ref{3B.5}) and inversely Laplace-transforming the result yields
\begin{eqnarray}
\frac{X(t_1)}{d} &=& \frac{i}{\eta T_0}(1\!-\!e^{-\eta t_1})
\!+\!\frac{u_1}{\eta^2}(e^{-\eta t_1}\!-\!1\!+\!\eta t_1)
\nonumber \\
& & +\!\int^\infty_0\!\frac{d\omega}{\pi}a_\omega(t)\!
\left[
\frac{i}{\omega}(1\!-\!e^{-\eta t_1})\!+\!\frac{i\eta}{\omega(\eta+i\omega)}
(e^{-\eta t_1}\!-\!e^{i\omega t_1})\right]\!.
\label{5A.2}
\end{eqnarray}
Substituting this into Eq.~(\ref{3B.6}) and solving for $a_\omega(t)$ gives
\begin{eqnarray}
a_\omega(t)\!&=&\!(1\!-\!e^{-i\omega t})e^{-i\omega T}\!+\!\frac{i}{\eta T_0}(f_\omega\!-\!g_\omega)\!+\!\frac{u_1}{\eta^2}(\eta j_\omega\!-\!f_\omega\!+\!g_\omega)\! \nonumber \\
& & + i\int^\infty_0\!\frac{d\omega'}{\pi}\frac{a_\omega'(t)}{\omega'}\!
\left[f_{\omega'}-g_{\omega'}+\frac{\eta}{\eta+i\omega'}(g_{\omega'}-h_{\omega'})\right],
\label{5A.3}
\end{eqnarray}
where
$f_\omega$, $g_\omega$, $h_\omega$, and $j_\omega$ are given by
\begin{eqnarray}
i\omega\int^T_0\!\!dt'\left[e^{-i\omega t'}+e^{-i\omega t}e^{i\omega (t'-2T)}
\right]k(t')
\label{5A.4}
\end{eqnarray}
with $k(t')=1, \ e^{-\eta t'}, \ e^{i\omega t'}$, and $t'$, respectively.

\subsection{Small-$\eta T_0$ expansion}

When $\eta T_0 \ll 1$, it is sufficient to keep terms
up to the first order in $\eta T_0$. Neglecting the higher-order terms, we obtain
\begin{eqnarray}
\frac{X(t_1)}{d}&=&\frac{i}{T_0}(t_1-\frac{\eta}{2}t_1^2)+\frac{u_1}{2}
t_1^2 +
\eta\int^\infty_0\!\frac{d\omega}{\pi}
\frac{a_\omega(t)}{\omega^2}(1\!+\!i\omega t_1\!-\!e^{i\omega t_1}),
\label{5B.1}
\end{eqnarray}
where
\begin{eqnarray}
a_\omega(t)=\frac{1}{\omega T_0}(1-e^{-i\omega T})(1-e^{-i\omega T}e^{-i\omega t}).
\label{5B.2}
\end{eqnarray}
The behavior of Eq.~(\ref{5B.1}) at large $t$ determines the transmission spectrum near zero bias, while its value at $t=0$ determines the total tunneling rate.
From condition $X(T)=d$ we find that
\begin{eqnarray}
\frac{T}{T_0}=-i\left[1+\frac{U_1T_0}{S_0}+\frac{\eta T_0}{2\pi}\right].
\label{5B.3}
\end{eqnarray}

We first consider the total tunneling rate which is determined by $S(t=0)$. 
We find that
\begin{eqnarray}
\frac{S(t=0)}{S_0}
=i\left[1-\frac{U_1T_0}{S_0}+\frac{\eta T_0}{\pi}\ln 2\right].
\label{5B.4}
\end{eqnarray}
The positive sign of the last term in the square bracket indicates that the total tunneling rate is suppressed by the interaction with the Ohmic bath~\cite{CL}.

We next consider the transmission spectrum near zero bias $E\sim E'$
which is dominated by the value of $S(t)$ at large $t$:
\begin{eqnarray}
\frac{S(t)}{S_0}\simeq i\left[1-\frac{U_1T_0}{S_0}
+\frac{\eta T_0}{4\pi}\left(3+i\pi+2\ln\frac{t}{T}\right)\right]\!.
\label{5B.5}
\end{eqnarray}
Substituting this into Eq.~(\ref{3A.1}) we find that
\begin{eqnarray}
\Gamma(E,E')\!&=&\!N
\frac{
\exp\left[-\frac{S_0}{\hbar}\left(1-\frac{U_1T_0}{S_0}
+\frac{3}{4\pi}\eta T_0\right)\right] 
}{
\Gamma\left(\frac{S_0}{2\pi\hbar}\eta T_0\right)
}
\frac{1}{|E-E'|}
\left[\frac{|E-E'|T_0}{\hbar}\right]^{\frac{S_0}{2\pi\hbar}\eta T_0}.
\label{5B.6}
\end{eqnarray}

\subsection{Zero-bias anomaly}
\label{sec:zero-bial}

Equation~(\ref{5B.6}) implies that the differential conductance obeys a power law near zero bias.
To be specific, we consider the case of electron tunneling between electrodes having the same chemical potential, and assumes for simplicity that the density of states, $D_{\rm L}, \ D_{\rm R}$, of the left and right electrodes are constant. 
The tunneling current $I(V)$ at bias voltage $V$ is then given by
\begin{eqnarray}
I(V)\!&=&\!eD_{\rm L}D_{\rm R}\int\!\!dE\!\int\!\!dE'\Gamma(E,E\!-\!E')
\left\{f(E)\left[1\!-\!f(E\!-\!E'\!+\!eV)\right]
\right.\nonumber \\
& & \left.-f(E)\left[1\!-\!f(E\!-\!E'\!-\!eV)\right]\right\}.
\label{5C.1}
\end{eqnarray}
At zero temperature, this reduces to
\begin{eqnarray}
& & I(V)=eD_{\rm L}D_{\rm R}\!\int^{eV}_0\!\!\!dE'\Gamma(E,E\!-\!E')(eV-E').
\label{5C.2}
\end{eqnarray}
Substituting Eq.~(\ref{5B.6}) into this yields
\begin{eqnarray}
I(V)\!&=&\!e^2D_{\rm L}D_{\rm R}N
\frac{
\exp\left[-\frac{S_0}{\hbar}\left(1-\frac{U_1T_0}{S_0}
+\frac{3}{4\pi}\eta T_0\right)\right] 
}{2+
\Gamma\left(\frac{S_0}{2\pi\hbar}\eta T_0\right)
}
V
\left[\frac{eVT_0}{\hbar}\right]^{\frac{S_0}{2\pi\hbar}\eta T_0}.
\label{5C.3}
\end{eqnarray}
Here $N$ is determined by the requirement that the differential tunneling resistance in the absence of the interaction at zero voltage is given by $R_T$:
\begin{eqnarray}
R_{\rm T}=\frac{dV}{dI}|_{V\rightarrow 0, \ \eta\rightarrow 0}.
\label{5C.4}
\end{eqnarray}
We thus find that
\begin{eqnarray}
I(V)\!=\!
\frac{
\exp\left[\frac{1}{\hbar}\!\left(eVT_0\!-\!\frac{3}{4\pi}\eta T_0S_0\right)\right] }{\Gamma\left(2+\frac{S_0}{2\pi\hbar}\eta T_0\right)
}
\frac{V}{R_{\rm T}}\!
\left[\frac{eVT_0}{\hbar}\right]^{\frac{S_0}{2\pi\hbar}\eta T_0}\!.
\label{5C.5}
\end{eqnarray}
This indicates that the differential conductance at low voltages $V$ obey a power law $\sim(eVT_0/\hbar)^{\eta T_0S_0/2\pi\hbar}$.


\figure{
Fig.~1.
Normalized tunneling exponent $S/S_0$ as a function of $\omega T_0$ for three values of $g$, where $S\equiv S_{\rm eff}(t=0)$, and $S_0$ and $T_0$ are the tunneling exponent and the barrier traversal time in the absence of interaction.
}
\figure{
Fig.~2.
Ratio of the first sideband intensities $I_{\pm}\equiv I_{\pm 1}$ plotted as a function of $\omega T_0$ for various values of $\beta\hbar\omega$, where $\beta^{-1}$ is the thermal energy and $\hbar\omega$ is the energy quantum of the single-frequency bath.
}


\begin{references}
\bibitem{Langer} J. S. Langer, Ann. Phys. (N.Y.) {\bf 41}, 108 (1967).
\bibitem{Coleman} S. Coleman, {\it Aspects of Symmetry}, (Cambridge University Press, Cambridge, 1985), p.~265.
\bibitem{Sebastian} K.~L.~Sebastian and G.~Doyen, Phys.~Rev.~B {\bf 47}, 7634 (1993), and references therein.
\bibitem{Bruinsma} R. Bruinsma and P. Bak, Phys.~Rev.~Lett. {\bf 56}, 420 (1986).
\bibitem{Grabert} For a collection of review articles, see, for example, {\it Single Charge Tunneling}, edited by H. Grabert and M. Devoret (Plenum, New York, 1991).
\bibitem{Mohring} K. M\"{o}hring and U. Smilansky, Nucl. Phys. {\bf A338}, 227 (1980).
\bibitem{Newns} D. M. Newns, Surf. Sci. {\bf 154}, 658 (1985).
\bibitem{Miller} W. H. Miller, in {\it Adv. Chem. Phys. XXV}, edited by
I. Prigogine and S. A. Rice (Wiley, New York, 1974); 
Acc. Chem. Res. {\bf 26}, 174 (1993).
\bibitem{Wolf} E. L. Wolf, {\it Principles of Electron Tunneling Spectroscopy}
(Oxford Science Publications, Oxford, 1985).
\bibitem{UA1} M. Ueda and T. Ando, Phys.~Rev.~B  {\bf 50}, 7820 (1994).
\bibitem{Feynman} R. P. Feynman and A. R. Hibbs, {\it Quantum Mechanics and Path Integrals}, (McGraw-Hill Inc., New York, 1965).
\bibitem{CL} A. O. Caldeira and A. J. Leggett, Ann. Phys. (N.Y.) 
{\bf 149}, 374 (1983).
\bibitem{BL} M. B\"uttiker and R. Landauer, Phys. Rev. Lett. {\bf 49}, 
1739 (1982); Phys. Scr. {\bf 32}, 429 (1985).
\bibitem{Landau} L. D. Landau and E. M. Lifshitz, {\it Quantum Mechanics}, (Pergamon, London, 1958).
\bibitem{Schwinger} J. Schwinger, J. Math. Phys. {\bf 2}, 407 (1961); 
K. V. Keldysh, Sov. Phys. JETP {\bf 20}, 1018 (1965) [Z. Eksp. Teor. Fiz. {\bf 47}, 1515 (1964)]; 
for a review, see, for example, K. C. Chou, Z. B. Su, B. L. Hau, and L. Yu, Phys.~Rep. {\bf 118}, 1 (1985).
\bibitem{Devoret} M. H. Devoret, D. Esteve, H. Grabert, G. -L. Ingold, 
H. Pothier, and C. Urbina, Phys. Rev. Lett {\bf 64}, 1824 (1990).
\bibitem{UA2} M.~Ueda and T.~Ando, Phys.~Rev.~B {\bf 52}, 16776 (1995).
\bibitem{Jonson} M. Jonson, Solid State Commun. {\bf 33}, 743 (1980).
\bibitem{Persson} B. N. J. Persson and A. Baratoff, Phys. Rev. B {\bf 38}, 9616 (1988). 
\bibitem{Gueret} P. Gu\'{e}ret, E. Marclay, and H. Meier, Solid State Commun. {\bf 68}, 977 (1988).
\bibitem{Esteve} D. Esteve, J. M. Martinis, C. Urbina, E. Turlot, and 
M. H. Devoret, Phys. Scr. {\bf T29}, 121 (1989).
\bibitem{UA3} M.~Ueda and T.~Ando, Phys.~Rev.~Lett. {\bf 72}, 1726 (1994);
ibid. {\bf 73}, 2785 (1994).
\bibitem{Nazarov} Yu. V. Nazarov, Zh. Eksp. Teor. Fiz. {\bf 95}, 975 (1989)
[Sov. Phys. JETP {\bf 68}, 561 (1989)].
\bibitem{Girvin} S. M. Girvin, L. I. Glazman, M. Jonson, D. R. Penn, and 
M. D. Stiles, Phys. Rev. Lett. {\bf 64}, 3183 (1990).
\end{references}
\end{document}